\documentclass[aps,prl,showpacs,twocolumn,superscriptaddress]{revtex4}
\usepackage[usenames]{color}
\newcommand{\comment}[1]{}
\usepackage{graphicx}
\usepackage{amsmath}
\usepackage{epsfig}
\usepackage{longtable}
\usepackage{float}

\begin{document}

\bibliographystyle{revtex}
\title{Understanding the nature of electronic effective mass in double-doped SrTiO$_{3}$}

\date{\today}
\author{J. Ravichandran} 
\affiliation{Applied Science and Technology Graduate Group,
University of California, Berkeley, Berkeley, CA 94720}
\author{W. Siemons}
\affiliation{Department of Physics, University of
California, Berkeley, CA 94720}
\author{M. L. Scullin}
\affiliation{Department of Materials Science and Engineering,
University of California, Berkeley, Berkeley, CA 94720}
\affiliation{Materials Sciences Division, Lawrence Berkeley National Laboratory, Berkeley, CA 94720}
\author{S. Mukerjee}
\affiliation{Department of Physics, University of California, Berkeley, CA 94720}
\affiliation{Department of Physics, Indian Institute of Science, Bangalore 560012, India}
\author{M. Huijben}
\affiliation{Department of Physics, University of California, Berkeley, CA 94720}
\affiliation{Faculty of Science and Technology and MESA+ Institute for Nanotechnology,
University of Twente, P.O. BOX 217, 7500 AE, Enschede, The Netherlands}
\author{J. E. Moore}
\affiliation{Department of Physics, University of California, Berkeley, CA 94720}
\affiliation{Materials Sciences Division, Lawrence Berkeley National Laboratory, Berkeley, CA 94720}
\author{R. Ramesh}
\affiliation{Department of Physics, University of California, Berkeley, CA 94720}
\affiliation{Department of Materials Science and Engineering, University of California, Berkeley, Berkeley, CA 94720}
\affiliation{Materials Sciences Division, Lawrence Berkeley National Laboratory, Berkeley, CA 94720}
\author{A. Majumdar}
\affiliation{Department of Materials Science and Engineering, University of California, Berkeley, Berkeley, CA 94720}
\affiliation{Materials Sciences Division, Lawrence Berkeley National Laboratory, Berkeley, CA 94720}
\affiliation{Department of Mechanical Engineering, University of California, Berkeley, Berkeley, CA 94720}
\affiliation{Applied Science and Technology Program, University of California, Berkeley, Berkeley, CA 94720}


\begin{abstract}
 We present an approach to tune the effective mass in an oxide semiconductor by a double doping mechanism. We demonstrate this in a model oxide system Sr$_{1-x}$La$_x$TiO$_{3-\delta}$, where we can tune the effective mass ranging from 6--20$\mathrm{m_e}$ as a function of filling or carrier concentration and the scattering mechanism, which are dependent on the chosen lanthanum and oxygen vacancy concentrations. The effective mass values were calculated from the Boltzmann transport equation using the measured transport properties of thin films of Sr$_{1-x}$La$_x$TiO$_{3-\delta}$. Our method, which shows that the effective mass decreases with carrier concentration, provides a means for understanding the nature of transport processes in oxides, which typically have large effective mass and low electron mobility, contrary to the tradional high mobility semiconductors.
\end{abstract}

\maketitle

 Effective mass is one of the fundamental quantities determining the transport properties of a material. Variation of effective mass as a function of carrier concentration and temperature has been widely characterized in small band-gap (0.1--1 eV) and low effective mass (0.1--0.01$\mathrm{m_e}$) semiconductors such as InAs~\cite{cardona}, HgTe~\cite{verie} and InSb~\cite{byszewski} and has been reviewed by Zawadzki~\cite{zawadzki}. Such systems show an increasing effective mass with carrier concentration which is sufficiently explained by Kane's band model~\cite{kane}. At the other end of the spectrum, complex oxides or transition metal oxides typically have larger effective masses (1--10$\mathrm{m_e}$) and band-gaps (1--10 eV). Complex oxides offer a variety of compounds showing no electron-electron correlation (band limit) to strong correlation (Mott limit). Strong correlations give rise to a very large effective mass (100--1000$\mathrm{m_e}$) in heavy fermionic systems~\cite{stewart} at cryogenic temperatures. Even though the effect of strong correlation on effective mass and other transport properties remains an intriguing question, there is little literature on the study of effective mass as a function of carrier concentration in oxides in the band limit. Moreover, a thorough understanding and tuneability of effective mass will have wide implications on phenomena such as thermoelectricity~\cite{okuda} and photovoltaics~\cite{green}. 

 In order to study the effective mass in the band limit, we have chosen SrTiO$_3$ (STO). STO is a model complex oxide system with very weak or no correlation and a wide range in n-type electrical conductivity, controlled by doping at the A-site (for example La doping in Sr sites), B-site (for example Nb doping in Ti sites) and by creating oxygen vacancies. The cation doping the A-site of STO is more suitable than doping on the B-site to study the nature of filling without changing the band structure drastically, because the conduction band has Ti 3$d$ characteristics. The effective mass of La doped STO has been reported as 6--6.6$\mathrm{m_e}$~\cite{sohta} and the introduction of oxygen vacancies results in a large effective mass of $\sim$16$\mathrm{m_e}$~\cite{frederikse} and has been attributed to the flat impurity band created by these vacancies~\cite{luo}. In this Letter, we demonstrate double doping in STO thin films by independently controlling oxygen vacancy concentration and La concentration. We performed optical spectroscopy to establish the formation of impurity band in STO due to oxygen vacancies. Transport measurements show the changes in scattering mechanism at the different limits of doping. We modeled our system using Boltzmann transport theory and show tuneability of the effective mass in the range of 6--20$\mathrm{m_e}$. 

Thin films (150 nm) of Sr$_{1-x}$La$_x$TiO$_{3-\delta}$ were grown via pulsed laser deposition (PLD) from dense polycrystalline, ceramic targets (each nominally containing either 0, 2 or 15\% La) onto (LaAlO$_3$)$_{0.3}$-(Sr$_2$AlTaO$_6$)$_{0.7}$ (LSAT) (001) single-crystal substrates ($a$=3.872 $\mathrm{\AA}$). Growth was carried out in oxygen partial pressures ranging from 10$^{-1}$--10$^{-7}$ Torr and a laser fluence of 1.75 J/cm$^2$ at a repetition rate of 8 Hz. Films were grown at a temperature of 450 $^\circ$C to create a non-equilibrium amount of oxygen vacancies, by avoiding the equilibrium reached during the cool down from higher temperatures. X-ray diffraction (XRD) was carried out on these films with a Panalytical X'Pert Pro thin film diffractometer using Cu K$\alpha$ radiation. Low temperature resistivity and Hall measurements were performed using a Quantum Design physical property measurement system (PPMS). Thermopower measurements at room temperature were done using a setup with T type thermocouples. UV-Visible (UV-Vis) transmission and reflection measurements were obtained from a Perkin Elmer Lambda 950 spectrometer and Hitachi U-3010 spectrometer respectively. The photoluminescence (PL) data was acquired using a setup with a 325 nm laser excitation source. 

\setlength{\epsfxsize}{1\columnwidth}
\begin{figure}[t]
\epsfbox{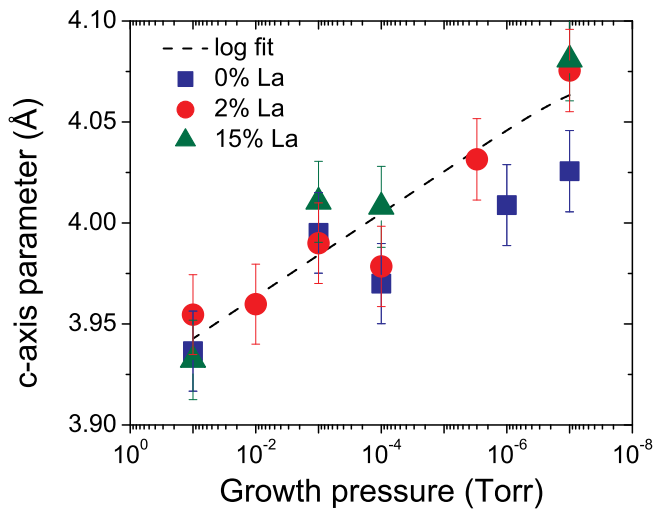}
\caption{The $c$-axis lattice parameter measured by x-ray diffraction at 300 K versus oxygen partial pressure during thin film growth for Sr$_{1-x}$La$_x$TiO$_{3-\delta}$ (the dashed line is a logarithmic fit to the 2\% data as a guide to the eye).} 
\label{Fig:latticeparam}
\end{figure}

XRD patterns of the thin films indicate that they are single phase, nearly single-crystal perovskites. The full-width-at-half-maximum (FWHM) of the rocking-curve (002) thin film peaks indicate the films are crystalline ($\omega_{002}^{\rm FWHM} < 0.3^\circ$) and (00$l$) epitaxially oriented. Typically, La doping has a negligible effect on the $c$-axis lattice parameter~\cite{sunstrom}, but oxygen vacancies will expand the lattice significantly as the Ti-Ti bond length is greater than that of Ti-O-Ti and induces a tetragonal distortion~\cite{luo,zhao}. Reciprocal space mapping of the (013) peak of the thin films reveal they are indeed tetragonal, with an in-plane lattice parameter of $\sim$3.90 $\mathrm{\AA}$ across all La concentrations and oxygen partial growth pressures, and a 0.023 $\mathrm{\AA}$ increase in the $c$-axis lattice parameter per order of magnitude decrease in oxygen partial pressure during growth (Fig.~\ref{Fig:latticeparam}). For samples grown at $10^{-7}$ Torr ---or a maximum concentration of oxygen vacancies in our study--- the $c$-axis lattice parameter is 4.075 $\mathrm{\AA}$ for Sr$_{0.98}$La$_{0.02}$TiO$_{3-\delta}$~\cite{footn1}, corresponding to a $c/a$ ratio of 1.045.

\setlength{\epsfxsize}{1\columnwidth}
\begin{figure}[t]
\epsfbox{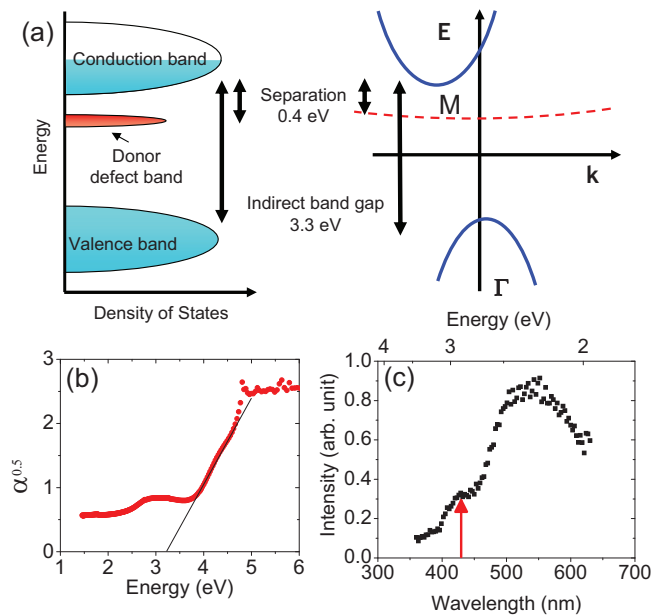}
\caption{(a) Schematic of the band model for the oxygen vacancy doped La-STO. (b) Plot of square root of absorbance as a function of photon energy for UV-Visible absorption spectroscopy. The intercept of the linear part of the curve gives the indirect band gap energy. (c) Plot of Photoluminescence intensity as a function of energy. The red arrow indicates the peak position corresponding to the energy gap between the impurity level and the valence band. The data shown corresponds to measurements at 300 K on 15\% La doped sample grown at $10^{-7}$ Torr.}
\label{Fig:bandmodel}
\end{figure}

This oxygen vacancy induced tetragonal distortion is expected to lift the three-fold $t_{2g}$ degeneracy of the conduction band. Theoretical predictions indicate that this leads to the formation of a heavy \emph{oxygen vacancy impurity band} lying below a light conduction band edge~\cite{luo,wunderlich,tanaka}. In order to validate the predicted band model, schematically shown in Fig.~\ref{Fig:bandmodel}, we performed UV-Vis spectroscopy and PL to map the important energy levels in the band model. UV-Vis spectroscopy was used to determine the energy gap between indirect band edge and the valence band and PL reveals the energy gap between the valence band edge and the oxygen vacancy impurity band. Fig.~\ref{Fig:bandmodel} shows the results obtained for 15\% La doped sample grown at $10^{-7}$ Torr. The indirect band gap determined from UV-Visible was $\sim$3.3 eV, very close to the value observed in bulk STO 3.27 eV~\cite{capizzi}. The PL spectra show a characteristic peak corresponding to $\sim$2.9 eV~\cite{mochizuki} suggesting a spacing of 0.4 eV between the indirect conduction band edge and the oxygen vacancy impurity band as predicted by theoretical calculations~\cite{luo,wunderlich}.

\setlength{\epsfxsize}{1\columnwidth}
\begin{figure}[t]
\epsfbox{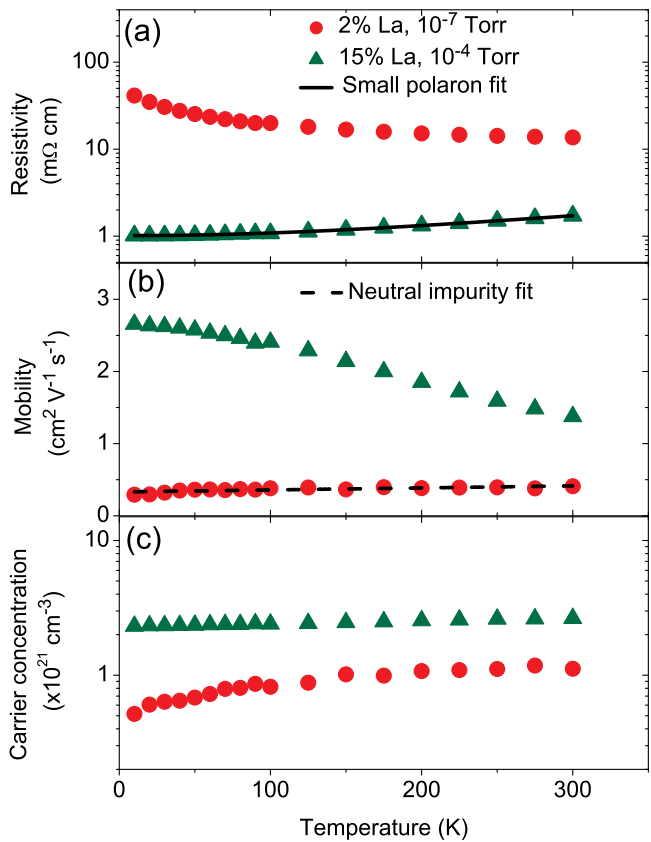}
\caption{Low temperature (a) resistivity (b) Hall mobility and (c) Hall carrier concentration data for 2\% La doped STO grown at $10^{-7}$ Torr and 15\% La doped STO grown at $10^{-4}$ Torr.}
\label{Fig:resistivity}
\end{figure}

Effective mass values have been conventionally evaluated using cyclotron resonance~\cite{dresselhaus}, reflectivity~\cite{vanmechelen}, photoemission~\cite{takizawa} and Shubnikov-de Haas effect~\cite{frederikse2} for various single crystals of semiconductors. The other common method is using transport data to calculate the density of states effective mass~\cite{frederikse,sohta}. Some of these methods are impractical for thin films and hence, we resorted to calculating the density of states effective mass using transport data. All transport data were measured in the plane of the (001)-oriented films. To evaluate the effective mass at room temperature, we need the low temperature mobility and resistivity data to learn about the dominant scattering mechanism and the carrier concentration and thermopower at room temperature. Typical low temperature resistivity, mobility and carrier concentration data are shown for two representative samples in Fig.~\ref{Fig:resistivity}. The measured transport parameters for all the samples at room temperature is recorded in Table \ref{table1}. The samples containing low doping of La (0\% or 2\%) but with oxygen vacancies showed a constant mobility as a function of temperature, indicating the presence of partially ionized impurity scattering or neutral impurity scattering~\cite{vining}. In this limit, the transport is dominated by the oxygen vacancies, which ionize partially if they are clustered together~\cite{shanthi,cuong} and clustering of vacancies in SrTiO$_3$ has been observed experimentally~\cite{muller}. In the case of samples containing 15\% La a small polaron conduction mechanism was observed by fitting the low temperature resistivity data to $\rho\sim{sinh}^{2}(\frac{1}{T})$. Recently, Liang \emph{et al.}~\cite{liang}, suggested the possibility of small polaron conduction in films grown under similar conditions. Fig.~\ref{Fig:resistivity} shows the fit for both the conduction mechanisms in the respective samples. The optical phonon mode's characteristic temperature ($T_{opt}$) derived by the fit was 100--120 K for the samples. It is interesting to note that this characteristic temperature is very close the soft mode transition in SrTiO$_3$ at 110 K~\cite{fleury}. This indicates that the suitable scattering parameter for these samples at 300 K is r=1 (for $T \ll T_{opt}$, r=$\frac{1}{2}$ and $T \gg T_{opt}$, r=1 ~\cite{askerov}).

 With the scattering mechanism known, we can solve the Boltzmann transport equations. To simplify our calculations, we have modeled our system with an effective single parabolic band with effective mass $m^*$. The value of $m^*$ were determined from the measurements of thermopower and carrier concentration $n$ (from the Hall mobility and resistivity). We have assumed that the presence of oxygen vacancies lifts the six fold degeneracy and hence the conduction band is only four fold degenerate. The equations used for the model ~\cite{vining} are

\begin{equation}
S = \mathrm{\frac{-k_B}{e}} \left[\frac{\left(r+2\right){F}_{r+1}\left(\eta\right)}{\left(r+1\right){F}_{r}\left(\eta\right)}-\eta\right]
\label{Eq:seebeck}
\end{equation} 
\begin{equation}
n = 2\pi z{\left(\frac{2{m}^{*}\mathrm{k_B}T}{{\mathrm{h}}^{2}}\right)}^{\frac{3}{2}}{F}_{\frac{1}{2}}\left(\eta\right)
\label{Eq:carrier}
\end{equation} 
\begin{equation}
  {F}_{r}\left(\eta\right)=\int_{0}^{\infty} \frac{x^{r}}{1+{e}^{x-\eta}}dx
\label{Eq:fermi}
\end{equation} 

where $\mathrm{k_B}$, h, e, $\eta$, $z$, $r$, ${m}^{*}$ are the Boltzmann constant, Planck's constant, electronic charge, reduced chemical potential $(\mu/\mathrm{k_B}T)$, degeneracy of the conduction band, scattering parameter and effective mass respectively. The scattering parameter gives the energy dependence of the scattering time and is of the form $\tau(\epsilon) = \tau_0 \epsilon^{r-\frac{1}{2}}$, where $\epsilon$ is the energy of the carrier. Knowing the scattering parameter from the temperature dependent mobility data, we can solve equations (\ref{Eq:seebeck}) and (\ref{Eq:carrier}), to obtain $\eta$ and ${m}^{*}$. All the measured and calculated values are listed in Table \ref{table1}.

\begin{center}
\begin{table*}[hb]
{\small
\caption{Various measured and derived physical quantities as a function of La doping and growth pressure in Sr$_{1-x}$La$_x$TiO$_{3-\delta}$ at 300 K; $\rho$ is resistivity, $S$ is thermopower, $n$ is Hall carrier density, $\mu$ is mobility, $r$ is scattering parameter, $m^{*}$ is effective mass, $\mathrm{m_e}$ is electron mass and $\eta$ is reduced chemical potential$(\mu/\mathrm{k_B}T)$.}
\begin{tabular*}{2\columnwidth}{@{\extracolsep{\fill}}cccccccccc}
\hline
\hline
S. No.&La\%&Growth pressure&$\rho$&$n$&$\mu$&$S$&$r$&$\frac{m^{*}}{\mathrm{m_e}}$&$\eta$\\
&&(Torr)&(m$\Omega$ cm)&(x${10}^{21}$ {cm}$^{-3}$)&cm$^2$V$^{-1}$s$^{-1}$&($\mu$V/K)&&&\\ 
\hline
1&0&${10}^{-7}$&16.3&0.6&0.64&-274&0.5&8.3&-0.6\\
2&2&${10}^{-3}$&1.5x${10}^{4}$&3.1x${10}^{-3}$&0.13&-832&0.5&18.6&-7.2\\
3&2&${10}^{-7}$&15.9&1.1&0.36&-190&0.5&7.1&0.6\\
4&15&${10}^{-3}$&4.6&2.1&0.65&-154&1&7.1&1.7\\
5&15&${10}^{-4}$&1.7&2.7&1.36&-101&1&6.1&2.7\\
6&15&${10}^{-7}$&2.7&3.1&0.75&-81&1&6.0&3.1\\
\hline
\hline
\label{table1}
\end{tabular*}}
\end{table*}
\end{center}

\setlength{\epsfxsize}{1\columnwidth}
\begin{figure}[ht]
\epsfbox{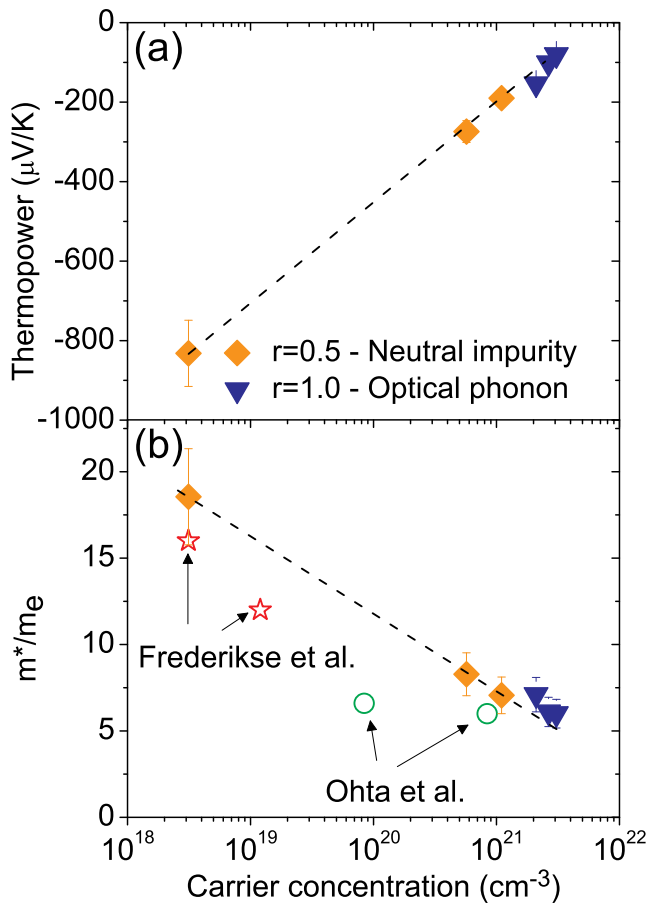}
\caption{(a) Measured thermopower and (b) calculated effective mass in Sr$_{1-x}$La$_x$TiO$_{3-\delta}$ as a function of carrier concentration $n$ at 300 K. Effective mass values from Ohta \emph{et al}.~\cite{sohta} and Frederikse \emph{et al}.~\cite{frederikse} are also shown for reference.}
\label{Fig:effmass}
\end{figure}

 The measured thermopower and the calculated values of the effective mass as a function of carrier concentration are plotted in Fig.~\ref{Fig:effmass}. The measured thermopower values for the samples at low carrier concentration ($\le {10}^{21}$ {cm}$^{-3}$) lie on log fit, indicating non-degenerate doping. For higher carrier concentration, we see some deviation from the logarithmic behavior as the doping tends to the degenerate limit. The reduced chemical potential listed in Table \ref{table1} also concurs with this line of analysis. Fig.~\ref{Fig:effmass}b shows we can influence the effective mass as a function of carrier concentration and scattering mechanism. Remarkably, the effective mass decreases with increasing carrier concentration unlike the conventional small band-gap semiconductors. This behavior is consistent with our assumption that the material should follow a simple parabolic model as has been observed in the past~\cite{sohta,okuda}. 

In summary, we have outlined a methodology to tune the effective mass in $n$-type oxide semiconductors through double doping with both an A-site dopant and oxygen vacancies in STO. The nature of $m^{*}$ is mainly filling controlled, specifically it decreases with increasing carrier concentration and can be tuned in the range of 6--20 $\mathrm{m_e}$ by chosing a given doping combination. The tuneability is achieved by the presence of a high effective mass impurity band which is separated from the conduction band by a small gap, as has been theorized prior to our observation of this energy level. A good understanding of the critical parameters may be used to better tailor the thermoelectric and photovoltaic response of oxide materials.  

The authors would like to acknowledge discussions with Choongho Yu, assistance of Dr. Costel Rotundu in Hall measurements, the UC Berkeley/LBNL thermoelectrics group and support from the US Dept. of Energy.


\begin{thebibliography}{23}
\bibitem{footn1}
\bibinfo{note}{Since the sensitivity of techniques such as EELS, RBS, and EDS is only $\sim$ 2\% for impurity atoms and much higher for vacancies, characterization of the number of oxygen vacancies can only be estimated through measurement of $n$, and the precise relationship between oxygen growth pressure and $\delta$ is unknown.}
\end{thebibliography}
\end{document}